\documentclass[twocolumn,aps,prl,showpacs,groupedaddress]{revtex4}

\usepackage{amsmath}
\usepackage{dcolumn}
\usepackage{epsfig}
\usepackage{graphicx}
\usepackage{latexsym}

\usepackage{epstopdf}

\begin{document}

\title{Suppression of Density Fluctuations in a Quantum Degenerate Fermi Gas}

\author{Christian Sanner}
\author{Edward J. Su}
\author{Aviv Keshet}
\author{Ralf Gommers}
\author{Yong-il Shin}
\author{Wujie Huang}
\author{Wolfgang Ketterle}
\affiliation{MIT-Harvard Center for Ultracold Atoms, Research
Laboratory of Electronics, and Department of Physics, Massachusetts
Institute of Technology, Cambridge MA 02139}

\begin{abstract}

We study density profiles of an ideal Fermi gas and observe Pauli
suppression of density fluctuations (atom shot noise) for cold clouds deep in the
quantum degenerate regime. Strong suppression is observed for
probe volumes containing more than 10,000 atoms. Measuring the level of suppression provides sensitive thermometry at low temperatures.
 After this method of sensitive noise measurements has been validated with an ideal
Fermi gas, it can now be applied to characterize phase transitions
in strongly correlated many-body systems.
\end{abstract}


\maketitle

Systems of fermions obey the Pauli exclusion principle. Processes
that would require two fermions to occupy the same quantum state are
suppressed. In recent years, several classic experiments have
directly observed manifestations of Pauli suppression in Fermi
gases.  Antibunching and the suppression of noise correlations are a direct consequence of the
forbidden double occupancy of a quantum state. Such experiments were
carried out for electrons~\cite{Yamamoto99ElectronHBT,Schonenberger99ElectronHBT,Hasselbach02ElectronHBT},
neutral atoms~\cite{Bloch06FermionAntibunching,Jeltes07HeliumAntibunching}, and neutrons~\cite{Pascazio06NeutronAntibunching}. In principle, such
experiments can be done with fermions at any temperature, but in
practice low temperatures increase the signal. A second class of (two-body)
Pauli suppression effects, the suppression of collisions,
requires a temperature low enough such that the de Broglie wavelength of the
fermions becomes larger than the range of the interatomic potential
and p-wave collisions freeze out. Experiments observed the
suppression of elastic collisions~\cite{Jin99PWaveThreshold,Jin01Collisions} and of clock shifts in
radio frequency spectroscopy~\cite{Ketterle03collisions,Ketterle03collisionsRFSpectroscopy}.

Here we report on the observation of Pauli suppression of density
fluctuations, a many-body phenomenon which occurs only at even lower temperatures in the quantum
degenerate regime, where the Fermi gas is cooled below the Fermi
temperature and the low lying quantum states are occupied with
probabilities close to one. In contrast, an ideal Bose gas close to quantum degeneracy shows
enhanced density fluctuations ~\cite{Bouchoule06EnhancedFluctuation}.

\begin{figure}[tbh!]
\begin{center}
\includegraphics[width=1.0\columnwidth]{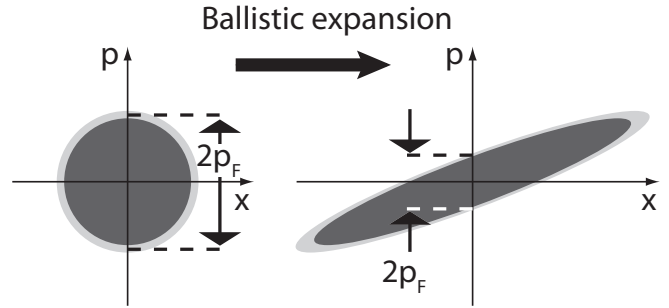}
\caption[]{Phase space diagram of ballistic expansion of a
harmonically trapped Fermi gas. Ballistic expansion conserves phase
space density and shears the initially occupied spherical area into
an ellipse.  In the center of the cloud, the local Fermi momentum and the sharpness of the Fermi
distribution are scaled by the same factor, keeping the ratio of
local temperature to Fermi energy constant. The same is true for all points in the expanded
cloud relative to their corresponding unscaled in-trap points.
\label{f:cartoon}}
\end{center}
\end{figure}

The development of a technique to sensitively measure density
fluctuations was motivated by the connection between density
fluctuations and  compressibility through the fluctuation
dissipation theorem. In this paper, we validate our technique for
determining the compressibility by applying it to the ideal Fermi gas. In
future work, it could be extended to interesting many-body
phases in optical lattices which are distinguished by their
incompressibility~\cite{Thivedi09Comppressibility}. These include the band insulator, Mott insulator,
and also the antiferromagnet for which spin fluctuations, i.e.
fluctuations of the difference in density between the two spin states,
are suppressed.

Until now, sub-Poissonian number fluctuations of ultracold atoms
have been observed only for small clouds of bosons with typically a
few hundred atoms ~\cite{Raizen05SubPoissonian,Oberthaler08SubPoissonian,Spreeuw10SubPoissonian,Steinhauer10SubPoissonian}
and directly~\cite{Chin09Incompressibility,GreinerUnpublished} or indirectly~\cite{Bloch02SubPoissonian}
for the bosonic Mott insulator in optical lattices.  For fermions in optical
lattices, the crossover to an incompressible Mott insulator phase 
was inferred from the fraction of double occupations~\cite{Esslinger05FermionCompressibility} or the
cloud size~\cite{Rosch08FermionCompressibility}.  Here we report the observation of density
fluctuations in a large cloud of fermions, showing sub-Poissonian
statistics for atom numbers in excess of 10,000 per probe
volume. 

\begin{figure}[tbh!]
\begin{center}
\includegraphics[width=1.0\columnwidth]{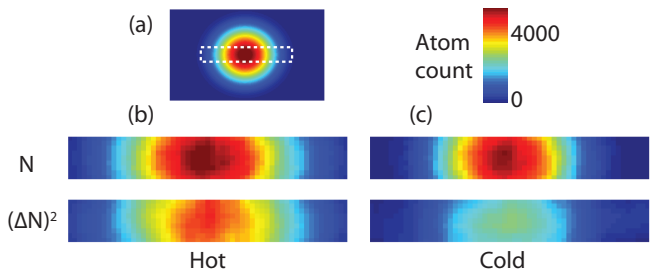}
\caption[]{(Color online) Comparison of density images to variance images. For
Poissonian fluctuations, the two images at a given temperature should be identical.  The
variance images were obtained by determining the local density
fluctuations from a set of 85 images taken under identical
conditions.  (a) Two dimensional image of optical density of an
ideal Fermi gas after 7 ms of ballistic expansion.  The noise data
were taken by limiting the field of view to the dashed region of
interest, allowing for faster image acquisition.  (b) For the heated
sample, variance and density pictures are almost identical, implying
only modest deviation from Poissonian statistics.  (c) Fermi
suppression of density fluctuations deep in the quantum degenerate
regime manifests itself through the difference between density and
variance picture.  Especially in the center of the cloud, there is a
large suppression of density fluctuations. The variance images were smoothed over 6$\times$6 bins. The width of images (b) and (c) is 2 mm. \label{f:false-color}} 
\end{center}
\end{figure}

The basic concept of the experiment is to repeatedly produce cold
gas clouds and then count the number of atoms in a small probe
volume within the extended cloud. Many iterations allow us to
determine the average atom number $N$ in the probe volume and its
variance $(\Delta N)^{2}$. For independent particles, one expects
Poisson statistics, i.e. $(\Delta N)^{2}/\langle N\rangle=1$. This is directly
obtained from the fluctuation dissipation theorem $(\Delta
N)^{2}/\langle N\rangle=n k_B T \kappa_T$, where $n$ is the density of the gas, and
$\kappa_T$ the isothermal compressibility. For an ideal classical gas
$\kappa_T=1/(n k_B T)$, and one retrieves Poissonian statistics. For
an ideal Fermi gas close to zero temperature with Fermi energy
$E_F$, $\kappa_T=3/(2 n E_F)$, and the variance $(\Delta N)^{2}$ is
suppressed below Poissonian fluctuations by the Pauli suppression
factor $3 k_B T/(2 E_F)$. All number fluctuations are thermal, as
indicated by the proportionality of $(\Delta N)^{2}$ to the
temperature in the fluctuation dissipation theorem. Only for the ideal
classical gas, where the compressibility diverges as $1/T$, one obtains
Poissonian fluctuations even at zero temperature.

The counting of atoms in a probe volume can be done while the atoms
are trapped, or after ballistic expansion. Ballistic expansion maintains the phase space density and
therefore the occupation statistics. Consequently, density
fluctuations are exactly rescaled in space by the ballistic
expansion factors as illustrated in Fig.\ref{f:cartoon}~\cite{Ketterle04Expansion,Charles00Rescale}. 
Note that this rescaling is a unique property of the harmonic oscillator potential, so future work on density fluctuations in optical lattices must employ in-trap imaging.  For the present work, we chose ballistic expansion.  This choice increases the number of fully resolved bins due to optical resolution and depth-of-field, it allows adjusting the optimum optical density by choosing an appropriate expansion time, and it avoids image artifacts at high magnification.

\begin{figure}[tbh!]
\begin{center}
\includegraphics[width=1.0\columnwidth]{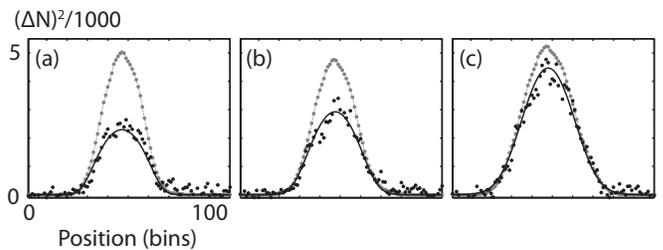}
\caption[]{Comparison of observed variances (black dots) with a theoretical model (black line) and the observed atom number (gray), at three different temperatures (a, b, and c),
showing 50, 40, and 15\% suppression. Noise thermometry is implemented by fitting the observed fluctuations, resulting in temperatures $T/T_F$ of 0.23$\pm$.01, 0.33$\pm$.02, and 0.60$\pm$.02. This is in good agreement with temperatures 0.21$\pm$.01, 0.31$\pm$.01, and 0.6$\pm$.1 obtained by fitting the shape of the expanded cloud ~\cite{Ketterle08Varenna}. The quoted uncertainties correspond to one standard deviation and are purely statistical. 
\label{f:noise-profile}}
\end{center}
\end{figure}

We first present our main results, and then discuss important
aspects of sample preparation, calibration of absorption cross
section, data analysis and corrections for photon shot noise.
Fig.~\ref{f:false-color}a shows an absorption image of an expanding
cloud of fermionic atoms. The probe volume, in which the number of
atoms is counted, is chosen to be 26 $\mu$m in the transverse directions, and extends through the
entire cloud in the direction of the line of sight. The large transverse size completely
avoids averaging of fluctuations due to finite optical resolution. From 85 such
images, after careful normalization~\cite{supp}, the variance
in the measured atom number is determined as a function of position.
After subtracting the photon shot noise contribution, a 2D image of
the atom number variance $(\Delta N)^{2}$ is obtained. For a
Poissonian sample (with no suppression of fluctuations), this image
would be identical to an absorption image showing the number of atoms per probe volume.
This is close to the situation for the hottest
cloud (the temperature was limited by the trap depth), whereas
the colder clouds show a distinct suppression of the atom
number variance, especially in the center of the cloud where the
local $T/T_F$ is smallest.

\begin{figure}[tbh!]
\begin{center}
\includegraphics[width=1.0\columnwidth]{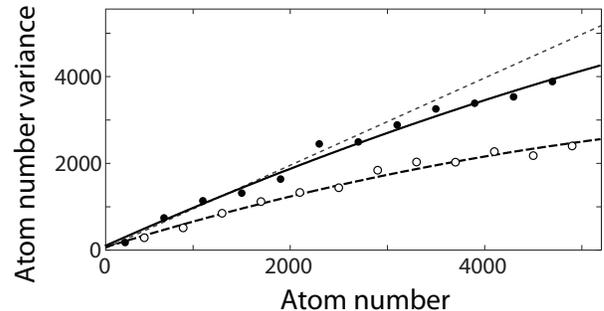}
\caption[]{Atom number variance vs. average atom number. For each spatial position, the average atom number per bin and its
variance were determined using 85 images. The filled and open circles in the figure are averages of different spatial bin positions with similar average atom number. 
For a hot cloud at $T/T_F$=0.6 (filled circles), the atom number 
variance is equal to the average atom number (dotted line, full Poissonian noise) in the spatial
wings where the atom number is low. The deviation from the linear
slope for a cold cloud at $T/T_F$=0.21 (open circles) 
is due to Pauli suppression of density fluctuations. There is also some suppression at the center of the hot cloud, where the atom number is high.
The solid and dashed lines are quadratic fits for the hot and cold clouds to guide the eye. 
 \label{f:var-plot}}
\end{center}
\end{figure}

In Fig.~\ref{f:noise-profile}, profiles of the variance are compared
to theoretical predictions~\cite{Castin08Varenna,PinesNozieres1998}.
Density fluctuations at wavevector q are
proportional to the structure factor $S(q,T)$.  Since our probe
volume (transverse size 26 $\mu$m) is much larger than the inverse
Fermi wavevector of the expanded cloud ($1/q_F=1.1 \mu$m), $S(q=0,T)$ has been integrated along the
line of sight for comparison with the experimental profiles. Within the local density approximation, $S(q=0,T)$ at a given position in the trap is the binomial variance $n_k(1-n_k)$ integrated over all momenta, where the occupation probability $n_k(k, \mu, T)$ is obtained from the Fermi-Dirac distribution with a local chemical potential $\mu$ determined by the shape of the trap. Fig.~\ref{f:var-plot} shows the dependence of the atom number variance on atom number for the hot and cold clouds. A statistical analysis of the data used in the figure is in ~\cite{supp}. 

The experiments were carried out with typically 2.5 million $^6$Li
atoms per spin state confined in a round crossed dipole trap with
radial and axial trap frequencies $\omega_{r}=2\pi\times160$~s$^{-1}$ and $\omega_{z}=2\pi\times230$~s$^{-1}$ corresponding to an in-trap
Fermi energy of $E_F = k_B\times2.15$~$\mu$K. The sample was prepared
by laser cooling followed by sympathetic cooling with $^{23}$Na in a
magnetic trap. $^6$Li atoms in the highest hyperfine state were transferred into
the optical trap, and an equal mixture of atoms in the lowest two hyperfine states
was produced. The sample was then evaporatively cooled
by ramping down the optical trapping potential at a magnetic bias
field B = 320$\pm$5 G where a scattering length of -300 Bohr radii
ensured efficient evaporation. Finally, the magnetic field
was increased to B = 520$\pm$5 G, near the zero crossing of the scattering
length, realizing a non-interacting Fermi gas. Absorption images
were taken after 7 ms of ballistic expansion. 

We were careful to prepare samples at different temperatures with similar cloud sizes and central optical densities to make sure that they were imaged with the same effective cross section and resolution. Hotter clouds were prepared by heating the colder cloud with parametric modulation of the trapping potential. For the hottest cloud this was done near 520 G to avoid excessive evaporation losses. 

Atomic shot noise dominates over photon shot noise only if each atom
absorbs several photons. As a result, the absorption images were
taken using the cycling transition to the lowest lying branch of the
$^2P_{3/2}$ manifold. However, the number of absorbed photons that
could be tolerated was severely limited by the acceleration of the
atoms by the photon recoil, which Doppler shifts the atoms out of
resonance. Consequently, the effective absorption cross section
depends on the probe laser intensity and duration. To limit the need
for nonlinear normalization procedures, we chose a probe laser
intensity corresponding to an average of only 6 absorbed photons per atom
during 4 $\mu$s of exposure time. At this intensity, about 12\% of
the $^6$Li saturation intensity, the measured optical density
 was found to be reduced by 20\% from its low-intensity
value~\cite{supp}. For each bin, the atom number variance is obtained by subtracting the known photon shot noise from the variance in the optical density ~\cite{supp}.

The absorption cross section is a crucial quantity in the
conversion rate between the optical density and the
number of detected atoms. For the cycling transition, the resonant
absorption cross section is $2.14\times 10^{-13}$ m$^2$. Applying the
measured 20\% reduction mentioned above leads to a value of $1.71\times 10^{-13}$ m$^2$. This is an upper limit to the cross section
due to imperfections in polarization and residual line broadening.
An independent estimate of the effective cross
section of $1.48\times 10^{-13}$ m$^2$ was obtained by comparing the
integrated optical density to the number of fermions necessary to
fill up the trap to the chemical potential. The value of the
chemical potential was obtained from fits to the ballistic expansion
pictures that allowed independent determination of the absolute
temperature and the fugacity of the gas. We could not precisely
assess the accuracy of this value of the cross section, since we did
not fully characterize the effect of a weak residual magnetic field
curvature on trapping and on the ballistic
expansion. The most accurate value for the effective cross section
was determined from the observed atom shot noise itself. The atom
shot noise in the wings of the hottest cloud is Poissonian, and this
condition determines the absorption cross section. Requiring that the slope of 
variance of the atom number $(\Delta N)^{2}$ vs. atom number $N$ is 
unity (see Fig.~\ref{f:var-plot}) results in a value of $(1.50\pm0.12)\times 
10^{-13}$ m$^2$ for the effective cross section in good
agreement with the two above estimates.

The spatial volume for the atom counting needs to be larger than the
optical resolution. For smaller bin sizes (i.e. small counting
volumes), the noise is reduced since the finite spatial resolution
and depth of field blur the absorption signal. In our setup, the smallest bin size without blurring
 was determined by the depth of field, since the size of the expanded
cloud was larger than the depth of field associated with the
diffraction limit of our optical system. We determined the effective
optical resolution by binning the absorption data over more and more
pixels of the CCD camera, and determining the normalized central
variance $(\Delta N)^2/N$ vs. bin size~\cite{supp}. The normalized variance increased and
saturated for bin sizes larger than 26 $\mu$m (in the object plane),
and this bin size was used in the data analysis. We observe the same suppression ratios for bin sizes as large as 40 $\mu$m, 
corresponding to more than 10,000 atoms per bin. 

For a cold fermion cloud, the zero temperature structure factor
$S(q)$ becomes unity for $q>2q_F$. This reflects the fact that
momentum transfer above $2q_F$ to any particle will not be Pauli
suppressed by occupation of the final state.  In principle, this can
be observed by using bin sizes smaller than the Fermi wavelength, or by 
Fourier transforming the spatial noise images. For
large values of $q$, Pauli suppression of density fluctuations
should disappear, and the noise should be Poissonian. However, our imaging system 
loses its contrast before $q\approx 2q_F$ ~\cite{supp}. 

Observation of density fluctuations, through the fluctuation-dissipation theorem,
determines the product of temperature and compressibility. It provides an absolute thermometer, as demonstrated in Fig.~\ref{f:noise-profile}
if the compressibility is known or is experimentally
determined from the shape of the density profile of the trapped
cloud~\cite{Ho09Comppressibility,Chin09Incompressibility}. Because variance is proportional to temperature for $T \ll T_F$, noise thermometry maintains 
its sensitivity at very low temperature, in contrast to the standard technique of fitting spatial profiles. 

Density fluctuations lead to Rayleigh scattering of light.
The  differential cross section for scattering light of wavevector $k$ by
an angle $\theta$ is proportional to the structure factor $S(q)$,
where $q=2k\sin(\theta/2)$~\cite{PinesNozieres1998}. In this work, we have
directly observed the Pauli suppression of density fluctuations and
therefore $S(q)<1$, which implies suppression of light scattering at
small angles (corresponding to values of $q$ inversely proportional
to our bin size).  How are the absorption images affected
by the suppression of light scattering?  Since the photon recoil was larger
that the Fermi momentum of the expanded cloud, large-angle light scattering is not suppressed. For the
parameters of our experiment, we estimate that the absorption cross
section at the center of a $T=0$ Fermi cloud is reduced by only 0.3\% due to Pauli
blocking~\cite{Ketterle01LightScattering}. Although we have not directly observed the Pauli suppression of light scattering, which 
has been discussed for over 20 years~\cite{Helmerson90LightScattering,Ketterle01LightScattering,Thywissen09LightScattering}, by observing reduced density fluctuations we have seen the underlying mechanism for suppression of light scattering. 

In conclusion, we have established a sensitive technique for determining
atomic shot noise and observed the suppression of density
fluctuations in a quantum degenerate ideal Fermi gas. This technique is promising for thermometry of
strongly correlated many-body systems and for observing
phase-transitions or cross-overs to incompressible quantum phases.

We acknowledge Joseph Thywissen and Markus Greiner for useful discussions. This work was
supported by NSF and the Office of Naval Research, AFOSR (through the Multidisciplinary
University Research Initiative program), and under Army Research Office grant
no. W911NF-07-1-0493 with funds from the Defense Advanced Research Projects
Agency Optical Lattice Emulator program.

\end{document}